\documentstyle[psfig,12pt,epsf]{article}
\begin{document}
\begin{titlepage}
\begin{centering}
\vspace*{1cm}
{\Large \bf Black Holes in the Framework of the}\\[4mm]
{\Large \bf Four-Dimensional Effective Theory}\\[3mm]
{\Large \bf of Heterotic Superstrings}\\[3mm]
{\Large \bf at Low Energies}\\
\vspace{2.5cm}
{\bf Panagiota Kanti}\footnote{\noindent Email~: pkanti@cc.uoi.gr . A full
postscript version of the dissertation in Greek (197 pages) can be retrieved
from the address~: http://artemis.sci.uoi.gr/\~{}pkanti/phd.ps.gz}\\[15mm]
{\it Physics Department, Division of Theoretical Physics,\\[2mm]
University of Ioannina, Greece}

\vspace*{0.5in}
\centerline{\thicklines \epsfysize=5cm \epsfbox{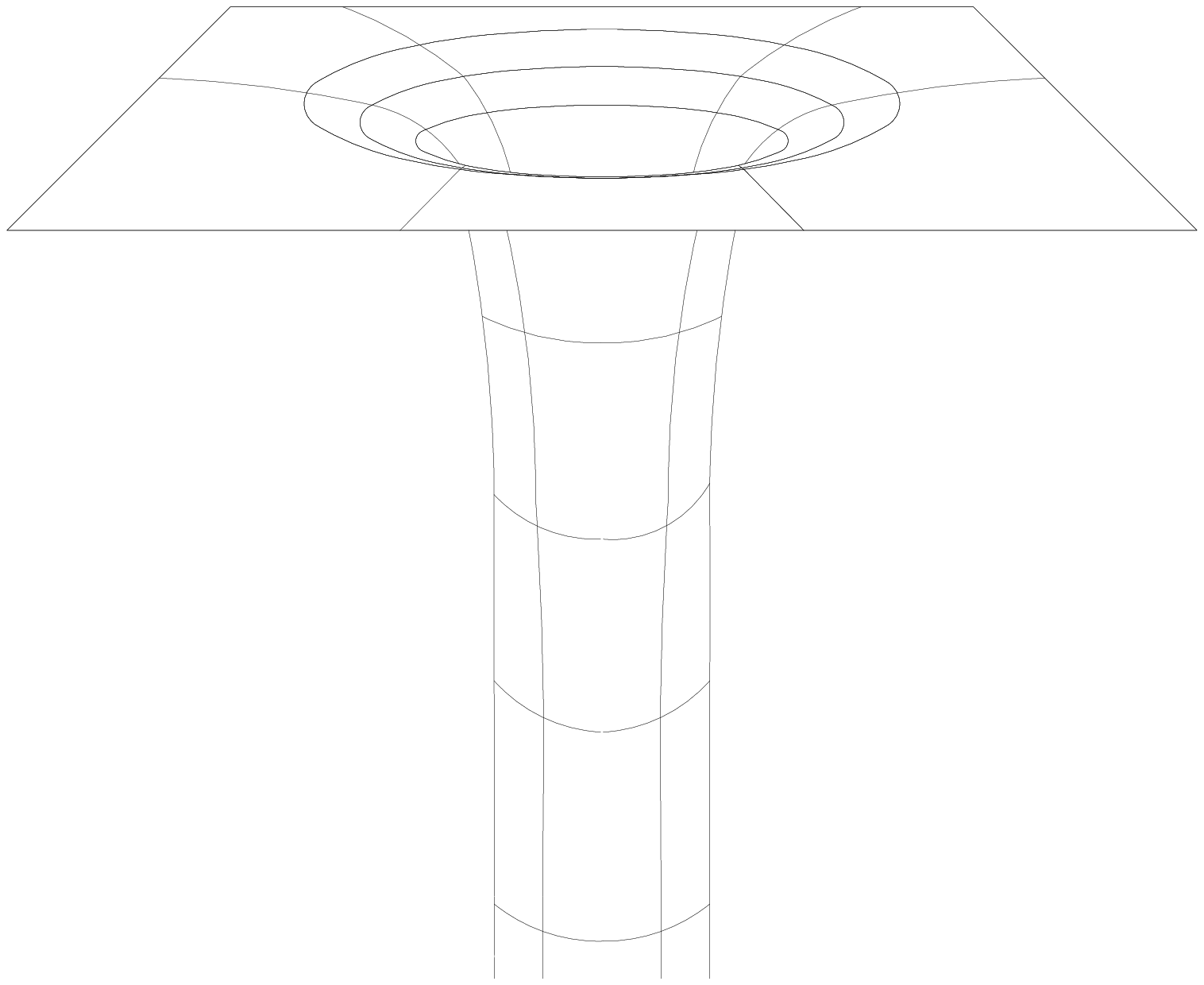}}

\vspace*{0.6in}
{\it A Dissertation in Candidacy for the Degree of Doctor of Philosophy}\\[5mm]
\centerline{{February 1998}}

\end{centering}
\end{titlepage}

\centerline{\Large \bf Summary}
\paragraph{}
Superstring Theory is the most promising unification scheme of the four forces
of nature. On the other hand, black holes are the first and most exciting
solutions of Einstein's equations and their properties are defined by various
theorems of the Theory of General Relativity, mainly by the ``No-hair"
theorem. The Low Energy Effective Superstring Theory plays the role of the
connecting link between these two theories and, at the same time, the role of
a generalised gravitational theory. In the framework of this effective theory,
we search for new black hole solutions. By extending existing treatments of
black hole solutions in string gravity, we compute the analy\-tical expressions
of all the scalar fields of the effective theory, that is of the dilaton, the
axions and the modulus field, outside the horizon of a Kerr-Newman black hole
and in ${\cal O}(\alpha')$~[1]. In the presence of higher curvature
gravitational terms, such as the quadratic Gauss-Bonnet term, we demonstrate
analytically the evasion of the existing ``no-hair" theorems including the
novel one by Bekenstein. In addition, we determine via numerical integration
the existence of a new family of black hole solutions, the so called Dilatonic
Black Holes, which are dressed with classical dilaton hair~[2]. In the
framework of the more general Einstein-Dilaton-Gauss-Bonnet-Yang-Mills theory,
another family of solutions, the Coloured Black Holes, are determined. These
solutions are characterized by non-trivial dilaton and Yang-Mills hair for the
particular case of $SU(2)$ gauge fields~[3]. The thermodynamical properties
of the black hole solutions are also discussed. The hair, however, found in all
of the above cases is of ``secondary type", in the sense that it owes its
existence to the primary gravitational or electromagnetic field. Finally, 
the linear stability of Dilatonic Black Holes under small spacetime-dependent
perturbations is exhibited through a semi-analytic method which makes use of
the Fubini-Sturm's theorem~[4]. This result is extremely important in that
it constitutes a linearly stable example of a black hole that bypasses the 
existing ``no-hair" theorems.\\[10mm]
\vspace*{5mm}
\centerline{\Large \bf Publications}
\noindent
[1] ``Classical Moduli ${\cal O}(\alpha')$ Hair", P. Kanti and K. Tamvakis,
Phys. Rev. {\bf D\,52} (1995)\\ \hspace*{0.65cm} 3506-3511.\\[2mm]
\noindent
[2] ``Dilatonic Black Holes in Higher Curvature String Gravity", P. Kanti,
N.E. Mavroma\-\hspace*{0.7cm}tos, J. Rizos, K. Tamvakis and E. Winstanley, 
Phys. Rev. {\bf D\,54} (1996) 5049-5058.\\[2mm]
[3] ``Coloured Black Holes in Higher Curvature String Gravity", P. Kanti and
K. Tamvakis, \\ \hspace*{0.65cm} Phys. Lett. {\bf B\,392} (1997) 30-38.\\[2mm]
[4] ``Dilatonic Black Holes in Higher Curvature String Gravity II~: Linear
Stability", \\ \hspace*{0.65cm} P. Kanti, N.E. Mavromatos, J. Rizos,
K. Tamvakis and E. Winstanley, Phys. Rev. \\ \hspace*{0.75cm}{\bf D\,57}
(to appear in the 15th May 1998 issue).

\newpage

\centerline{\Large\bf CONTENTS}
\vspace*{8mm}
\noindent
{Preface} \dotfill 1\\[5mm]
{\bf 1. Introduction to String Theory} \dotfill 3\\[3mm]
\indent 1.1 Introduction \dotfill {\rm 3}\\[2mm]
\indent 1.2 Bosonic String \dotfill {\rm 4}\\[2mm]
\indent 1.3 Superstring \dotfill {\rm 16}\\[2mm]
\indent 1.4 Heterotic Superstring and Compactification \dotfill {\rm 29}\\[2mm]
\indent 1.5 Effective String Theory at Low Energies \dotfill {\rm 36}\\[2mm]
\indent 1.6 Four-Dimensional Effective Theory and Moduli Fields \dotfill
{\rm 45}\\[5mm]
{\bf 2. Introduction to Black Holes} \dotfill {\rm 49}\\[3mm]
\indent 2.1 Introduction \dotfill {\rm 49}\\[2mm]
\indent 2.2 The Schwarzschild Metric\dotfill {\rm 52}\\[2mm]
\indent 2.3 The Kerr-Newman Metric \dotfill {\rm 60}\\[2mm]
\indent 2.4 Killing Vectors and Surface Gravity \dotfill {\rm 64}\\[2mm]
\indent 2.5 Thermodynamics of Black Holes \dotfill {\rm 66}\\[2mm]
\indent 2.6 The ``No-hair" Theorem \dotfill {\rm 73}\\[5mm]
{\bf 3. ``Scalar Hair" in Kerr-Newman Black Hole Background} \dotfill
{\rm 81}\\[3mm]
\indent  3.1 Introduction \dotfill {\rm 81}\\[2mm]
\indent  3.2 Action Functional and Equations of Motion of the Theory
\dotfill {\rm 86}\\[2mm]
\indent  3.3 Solution at Zero Order in $\alpha'$ \dotfill {\rm 88}\\[2mm]
\indent  3.4 Solution at First Order in $\alpha'$ \dotfill {\rm 90}\\[2mm]
\indent  Appendix 3.A Derivation of the Equations of Motion \dotfill
{\rm 97}\\[2mm]
\indent  Appendix 3.B The Kerr Metric \dotfill {\rm 103}\\[5mm]
{\bf 4. Black Holes with ``Dilaton Hair" in the Presence of Higher Curvature 
\\[1mm] \indent Gravitational Terms} \dotfill {\rm 105}\\[3mm]
\indent  4.1 Introduction \dotfill {\rm 105}\\[2mm]
\indent  4.2 Action Functional and Equations of Motion of the Theory
\dotfill {\rm 106}\\[2mm]
\indent  4.3 The Novel ``No-hair" Theorem \dotfill {\rm 111}\\[2mm]
\indent  4.4 Numerical Analysis \dotfill {\rm 115}\\[2mm]
\indent \indent 4.4.1 Determination of Black Hole Solutions \dotfill
{\rm 115}\\[2mm]
\indent \indent 4.4.2 Determination of the Independent Parameters of the Black 
Hole \\[1mm] \indent \indent \hspace*{0.9cm} Solutions at Infinity \dotfill
{\rm 119}\\[2mm]
\indent \indent 4.4.3 Determination of Additional Solutions \dotfill
{\rm 122}\\[2mm]
\indent Appendix 4.A The Coefficients $d$, $d_1$ and $d_2$\dotfill
{\rm 127}\\[5mm]
\noindent
{\bf 5.  ``Coloured" Black Holes in the Presence of Higher Curvature
\\[1mm] \indent Gravitational Terms} \dotfill {\rm 129}\\[2mm]
\indent  5.1 Introduction \dotfill {\rm 129}\\[2mm]
\indent  5.2 Action Functional and Equations of Motion of the Theory
\dotfill {\rm 130}\\[2mm]
\indent 5.3 Asymptotic Form of the Solutions at Infinity and the Event Horizon 
\dotfill {\rm 133}\\[2mm]
\indent  5.4   Thermodynamical Study of Black Hole Solutions \dotfill
{\rm 137}\\[2mm]
\indent  5.5  Numerical Analysis \dotfill {\rm 142}\\[2mm]
\indent  \indent 5.5.1  Determination of Black Hole Solutions \dotfill
{\rm 143}\\[2mm]
\indent  \indent 5.5.2  Determination of ``Naked Singularities" 
\dotfill {\rm 150}\\[5mm]
{\bf 6. Stability Analysis of Black Holes with ``Dilaton ``Hair"}
\dotfill {\rm 153}\\[2mm]
\indent  6.1 Introduction \dotfill {\rm 153}\\[2mm]
\indent  6.2 Equations of Motion of the Time-Dependent Theory 
\dotfill {\rm 154}\\[2mm]
\indent  6.3 Linear Stability Analysis \dotfill {\rm 157}\\[2mm]
\indent  6.4 Asymptotic Behaviour of Perturbation $\delta\phi$ 
\dotfill {\rm 164}\\[2mm]
\indent  6.5 Application of the Fubini-Sturm's Theorem \dotfill {\rm 167}\\[2mm]
\indent  6.6 Stability of the ``Coloured" Black Holes 
\dotfill {\rm 172}\\[5mm]
\bf 7. Conclusions \dotfill {\rm 177}\\[5mm]
\bf Bibliography \dotfill {\rm 185}

\end{document}